# Magneto-optical and micromagnetic simulation study the current driven domain wall motion in ferromagnetic (Ga,Mn)As

K. Y. Wang[1], A. C. Irvine[2], R. P. Campion[3], C. T. Foxon[3], J. Wunderlich[1], D. A. Williams[1], and B. L. Gallagher[3]

[1] *Hitachi Cambridge Laboratory, Cambridge CB3 0HE, United Kingdom*

[2] *Microelectronics Research Centre, Cavendish Laboratory, University of Cambridge, CB3 0HE, United Kingdom*

[3] *School of Physics and Astronomy, University of Nottingham, Nottingham NG7 2RD, United Kingdom*

We have studied current-driven domain wall motion in modified $Ga_{0.95}Mn_{0.05}As$ Hall bar structures with perpendicular anisotropy by using spatially resolved Polar Magneto-Optical Kerr Effect Microscopy and micromagnetic simulation. Regardless of the initial magnetic configuration, the domain wall propagates in the opposite direction to the current with critical current of $1\sim2\times10^5 A/cm^2$. Considering the spin transfer torque term as well as various effective magnetic field terms, the micromagnetic simulation results are consistent with the experimental results. Our simulated and experimental results suggest that the spin-torque rather than Oersted field is the reason for current driven domain wall motion in this material.

*Keywords: spin-torque; magneto-optical; micromagnetic*

# 1: Introduction:

Manipulation of magnetization and the magnetic domain wall motion using spin polarized current has attracted much attention for potential applications in spintronics, such as novel magnetic devices using written information with an electrical current. The spin polarized current gives the angular momentum to the local magnetization, which is called the spin-transfer torque [1]. Current driven domain wall motion in both the ferromagnetic metals and magnetic semiconductors has been demonstrated, although the mechanism is still in debate [2-4]. The critical current density for domain wall motion is predicted to be proportional to the saturation magnetization [5,6]. In diluted magnetic semiconductor (Ga,Mn)As doped with Mn 1-10%, the saturation magnetization is one to two orders of magnitude smaller than that of the metallic ferromagnets ( The experimentally observed critical current for ferromagnetic semiconductor (Ga,Mn)As is 1 to 2 orders of magnitude smaller than that in ferromagnetic metals[7]). The lower electrical current, the heating effect and the Oersted field produced by the electrical current are weaker. Because of the low critical current density, a ferromagnetic semiconductor is one of the best candidates for understanding current driven domain wall motion. In this letter, we use Polar Magneto-Optical Kerr Microscopy (PMOKM) and LLG micromagnetic simulations [8] to study the current driven domain wall motion in $Ga_{0.95}Mn_{0.05}As$ modified Hall bar devices.

# 2: Experimental Results

25 nm (Ga,Mn)As epilayers grown on a relaxed (001) (In,Ga)As buffer layer experience a tensile strain due to the difference in the lattice constant in each layer. Under these growth conditions, the magnetic easy axis is known to be perpendicular to the growth plane in agreement with the theory. Magnetic domain patterns in the (Ga,Mn)As thin films have been studied previously using PMOKM [9]. During the magnetic reversal, in-situ imaged domain patterns clearly show that the domain walls are nucleated and propagate along the film. It suggests that we can easily make use of the nucleated domain walls for studying current driven domain wall motion in these materials. A 4μm wide modified Hall bar device was fabricated with electron beam lithography, as shown in figure 1. The length between the two neighbour arms along the bar is 20μm and the total length of the bar is 120 μm. A 10nm surface layer has been etched away at the both ends, as shown in figure 1. Due to the different coercive field for the etched and non-etched part of the devices, we initialize two domain walls at both interfaces by an external magnetic field. No matter what the relation between the applied current direction and domain wall motion direction, we expect to see the domain wall motion under the Kerr microscopy image window.

Initially, the device is saturated with a negative magnetic field H = -300 Oe, which is much higher than the coercive field. The field is then swept to H=45 Oe, which only switches part of the device with the two domain walls formed at the interface. The image of the initial magnetic configuration is shown in figure 2 a. During the application of the dc current, we screen the light in order to prevent the complication of light excited effects. Increasing dc current from zero to negative value at a rate of 2 $\times 10^4$ A/cm$^2$.s, we monitor the Hall resistance for both the A and B arm pairs simultaneously. Large anisotropic magnetoresistance and anomalous Hall Effect was observed in GaMnAs devices because of strong spin-orbit coupling[10]. The anomalous Hall Effect is very sensitive to the local magnetization. When we observe an abruptly change of the Hall resistance, we switch the dc current to zero. The monitored Hall resistance for A and B arm pairs are shown in figure 2 B. The longitudinal resistance increases by about 3% during the applied dc current. Compared with the temperature dependent of the longitudinal resistance measured at low current density j = $1 \times 10^3$ A/cm$^2$, the temperature rise because of heating during the applying dc current is less than 10K. The magnetic configuration image after applied the dc current of figure 2b is shown in figure 2 c. The right side domain wall propagates about 60 μm along the channel to the left side. The direction of motion is opposite to the electrical current direction with critical current ~ 1-2 $\times 10^5$ A/cm$^2$. Regardless of the initial magnetic configuration, we find that the domain wall motion direction is always opposite to the electrical current direction, which is in agreement with other experiments [4,7]. The heating effect has no preferred direction. If the Current generated Oersted field is the origin of the domain wall motion, the domain wall motion direction should be sensitive to the initial magnetic configuration. It shows that the spin transfer torque rather than the heating effect and the Oersted field is the origin of the current driving domain wall motion. The relation between the current direction and domain wall motion can be explained by the fact that because (Ga,Mn)As is a p type conductor, the electrical current and the carriers move in the same direction. The interaction between the itinerate holes and localized Mn spins is p-d antiferromagnetic coupling [11]. The spin-polarized holes give the momentum to the localized magnetization. The domain wall motion direction is expected to be opposite to the current direction, which is in agreement with the observed results.

### 3: Micromagnetic simulations:

In this section, the micromagnetic simulations were carried out to study the current driven domain wall in a GaMnAs slab with the physical dimensions (5200nm length, 640 nm wide and 20nm thick). The Gilbert damping constant $\alpha = 0.02$ [12], the magnetic anisotropy constant $K_U = 2 \times 10^3 J/m^3$ and saturation magnetization $M_S = 25 emu/cm^2$ are used. The initial magnetic state with one stable domain wall is shown in the figure 3a. We have simulated the current driven domain wall motion at different current densities with other simulated parameters fixed. The magnetic configuration after applying 6ns dc current is shown in figure 3 b-d. The critical current density for the domain wall motion is $2.5 \times 10^5$ A/cm$^2$, which is only 1 to 2 times larger than that of the experimental results. The domain wall displacement increases with increasing applied current density. The domain wall is still quite sharp and clean at low current density and distorted by the Oersted field at high current density. The distortion of the domain wall from Oersted field is underestimated here, because we have not considered the Oersted field generated by the anomalous Hall field [7]. The simulated results also show convincingly that the spin-torque is the origin of the current driving domain wall motion. In the simulation, we have not considered the non-adiabatic term and the heating effect from the electrical current for the current driving domain wall. We will study this in more detail elsewhere.

## Conclusions:

Combining experimental MOKM measurements and micromagnetic simulations, we have studied the current driven domain wall motion in modified $Ga_{0.95}Mn_{0.05}As$ Hall bar devices. The domain wall always moves in opposite direction to the current independent of the initial magnetic configuration. The critical current density obtained from the experiments is $1\sim2 \times10^5 A/cm^2$, which is in agreement with the simulated results.

*Acknowledgements:*

This project was supported by EC sixth framework Grant No. FP6-IST-015728 and EPSRC-GB(GR/S81407/01).

*Figure Captions:*

*Fig.1: The fabricated modified multi-arms Hall bar device with width 4 μm and the length between neighbour arms is 20 μm. The surface of the two square marked parts has been etched away about 10 nm.*

Fig.2: *(a) Initial magnetic configuration with domain wall at both sides. (b) During applying dc current, the in-situ monitored Hall resistance at bar A and B; (c) Final magnetic configuration after applying dc current. The domain wall motion direction is opposite to the electrical current direction.*

*Fig.3: Micromagnetic simulation results: (a) Initial magnetic configuration; magnetic configuration after applying dc current 6 ns with current density (b) $J=2.5\times10^5 A/cm^2$; (c) $J=2\times10^6 A/cm^2$ and (d) $J=1\times10^7 A/cm^2$. Arrow points the electrical current direction.*

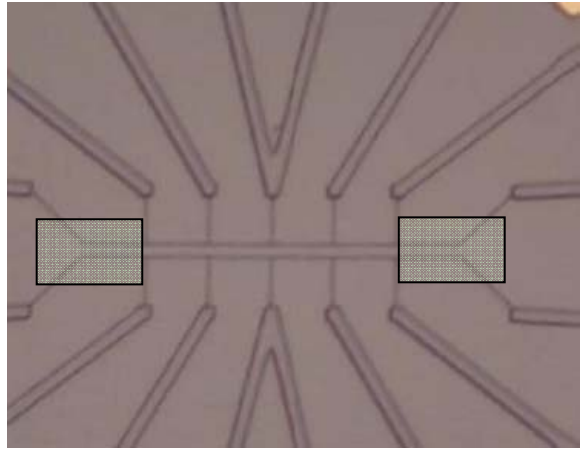

*Figure 1 K. Y. Wang et al.*

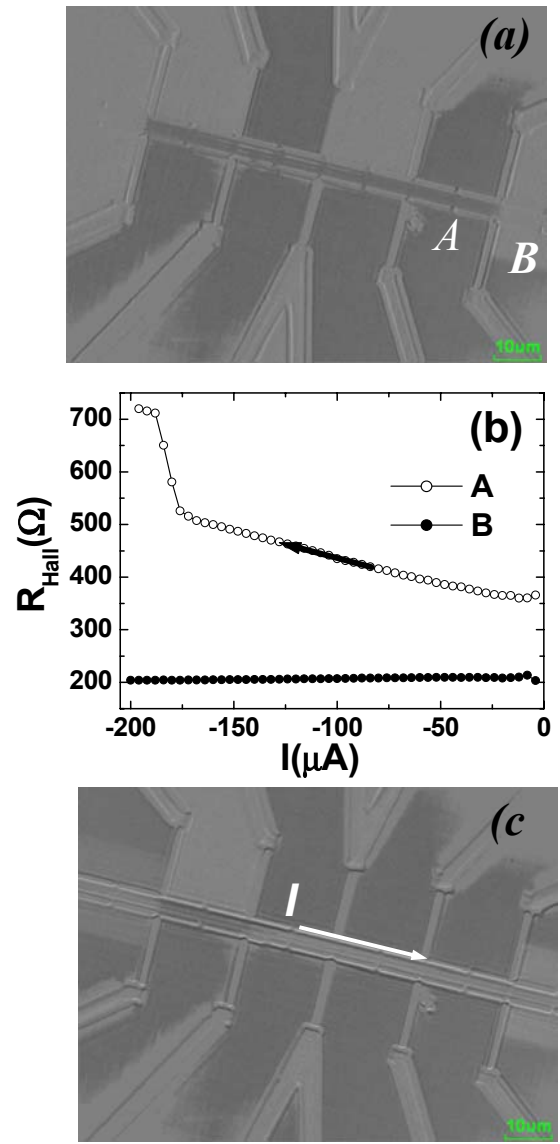

Figure 2 K. Y. Wang et al.

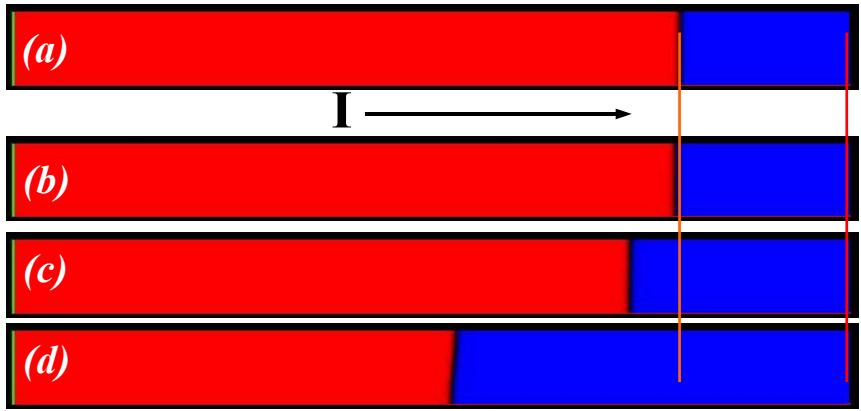

*Figure 3 : K. Y. Wang et al.*